\documentclass[aps,pre,twocolumn,superscriptaddress,longbibliography]{revtex4-1}
\usepackage{multirow}
\usepackage{dcolumn}
\usepackage{epsfig}
\usepackage{xcolor}
\usepackage{graphicx}
\usepackage{dcolumn}
\usepackage{bm}
\usepackage{amsmath}
\usepackage{hyperref}
\newcommand{\PreserveBackslash}[1]{\let\temp=\\#1\let\\=\temp}
\newcolumntype{C}[1]{>{\PreserveBackslash\centering}p{#1}}
\newcolumntype{R}[1]{>{\PreserveBackslash\raggedleft}p{#1}}
\newcolumntype{L}[1]{>{\PreserveBackslash\raggedright}p{#1}}

\begin{document}

\title{Machine learning prediction of network dynamics with privacy protection}

\author{Xin Xia}
\affiliation{The Key Laboratory of Intelligent Computing and Signal Processing of Ministry of Education, School of Mathematical Science, Anhui University, Hefei 230601, China}

\author{Yansen Su}
\affiliation{The Key Laboratory of Intelligent Computing and Signal Processing of Ministry of Education, School of Artificial Intelligence, Anhui University, Hefei 230601, China}

\author{Linyuan L\"{u}}
\affiliation{Institute of Fundamental and Frontier Sciences, University of Electronic Science and Technology of China, Chengdu, 610054, China}

\author{Xingyi Zhang}
\affiliation{The Key Laboratory of Intelligent Computing and Signal Processing of Ministry of Education, School of Artificial Intelligence, Anhui University, Hefei 230601, China}

\author{Ying-Cheng Lai}
\affiliation{School of Electrical, Computer and Energy Engineering, Arizona State University, Tempe, Arizona 85287, USA}

\author{Hai-Feng Zhang} \email{haifengzhang1978@gmail.com}
\affiliation{The Key Laboratory of Intelligent Computing and Signal Processing of Ministry of Education, School of Mathematical Science, Anhui University, Hefei 230601, China}

\date{\today}

\begin{abstract}

	Predicting network dynamics based on data, a problem with broad applications, has been studied extensively in the past, but most existing approaches assume that the complete set of historical data from the whole network is available. This requirement presents a great challenge in applications, especially for large, distributed networks in the real world, where data collection is accomplished by many clients in a parallel fashion. Often, each client only has the time series data from a partial set of nodes and the client has access to only partial timestamps of the whole time series data and partial structure of the network. Due to privacy concerns or license related issues, the data collected by different clients cannot be shared. To accurately predict the network dynamics while protecting the privacy of different parties is a critical problem in the modern time. Here, we propose a solution based on federated graph neural networks (FGNNs) that enables the training of a global dynamic model for all parties without data sharing. We validate the working of our FGNN framework through two types of simulations to predict a variety of network dynamics (four discrete and three continuous dynamics). As a significant real-world application, we demonstrate successful prediction of State-wise influenza spreading in the USA. Our FGNN scheme represents a general framework to predict diverse network dynamics through collaborative fusing of the data from different parties without disclosing their privacy.

\end{abstract}
\maketitle

\section{Introduction} \label{sec:intro}

This paper deals with the problem of predicting complex network dynamics from
distributed data without compromising the privacy of the data sources. In
particular, given a large network whose dynamics are unknown but only
{\em local} historical data or time series are available through measurements
conducted by different parties (clients or agents), the objective is to
accurately predict the dynamical evolution of the network for a number of time
steps under the constraint of no data sharing of any kind among the clients. To
paraphrase it, among the clients who performed the measurements, there can be
no communication of any sort for privacy considerations. This problem of
predicting network dynamics without privacy disclosure is significantly
more challenging than previously studied inverse problems in the field of
reverse engineering of networked dynamical systems. The main contribution of
this paper is the articulation and validation of an effective machine learning
based solution to this problem.

To infer or reconstruct the dynamical process on a network based on time series
data has been an active field in the past two decades~\cite{WLG:2016,wang2022full}.
A diverse array of methodologies were proposed, including those based on the
collective dynamics~\cite{YRK:2006,ZXZXC:2013,Timme:2007,TC:2014,NCT:2017},
stochastic analysis~\cite{MZL:2007,LL:2017}, compressive
sensing~\cite{WYLKG:2011,WYLKH:2011,SWFDL:2014,SWL:2014,MWWHLC:2018,Lai:2021},
and machine learning~\cite{lin2011machine,KFGL:2021a,KFGL:2021b,kumar2021applications}.
In nonlinear dynamics, the research on data based identification and forecasting
of system dynamics has an even longer history~\cite{FS:1987,Casdagli:1989}.
For example, an earlier approach focused on approximating a nonlinear system
by various linear equations in different regions of the phase space so that the
local Jacobian matrices can be
constructed~\cite{FS:1987,Gouesbet:1991,Sauer:1994} or the ordinary
differential equations can be found to fit the data~\cite{BBBB:1992}.
Methods based on chaotic synchronization~\cite{Parlitz:1996}
for estimating the system parameters were also investigated.
Of particular importance is the approach to finding the system equations
(hence the system dynamics) from data. This ``natural'' approach dated back
to the original work of Crutchfield and McNamara~\cite{CM:1987}, who exploited
the concept of qualitative information to deduce the effective equations of
motion of the system responsible for the deterministic portion of the observed
random behavior. An inverse Frobenius–Perron approach to
generate a dynamical system close to the original system in the sense of the
invariant density was proposed~\cite{Bollt:2000}. The Kronecker product
representation was also used for modeling and nonlinear parameter
estimation~\cite{YB:2007}. In the past decade, sparse optimization methods,
e.g., compressive sensing~\cite{CRT:2006a,CRT:2006b},
was introduced for finding the system equations from
data~\cite{WYLKG:2011,WLGY:2011,WYLKH:2011,SWFDL:2014,Lai:2021}
for nonlinear and complex dynamical systems whose velocity fields or mapping
functions are describable by a number of fairly elementary mathematical
functions. This equation-finding approach, while appealing and satisfying from
a mathematical point of view, may not have significant practical value in real
world applications, as the dynamical processes there often cannot be described
by a collection of simple functions. Even in cases where an approximate set of
equations can be found, sensitivity to small errors typically seen in nonlinear
and complex dynamical systems can lead to large deviations between the dynamics
as predicted by the equations and the ground truth. In these situations,
machine learning has gained recent attention as a viable approach to predicting
dynamics from data~\cite{Raissi:2018,PHGLO:2018,PWFCHGO:2018,GPG:2019,QWX:2019,ZZLTXZ:2019,FJZWL:2020,ZJQL:2020,ZW:2020,KFGL:2021a,KFGL:2021b,KLNPB:2021}.

\begin{figure} [ht!]
\centering
\includegraphics[width=\linewidth]{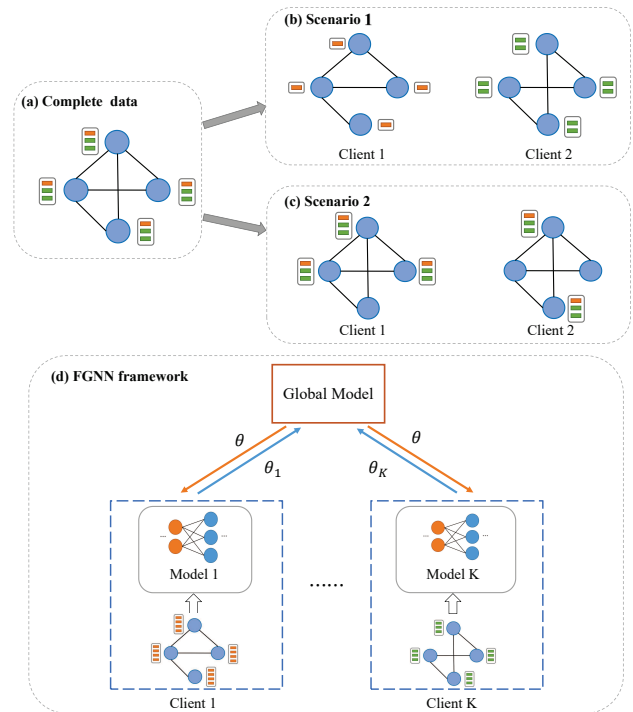}
\caption{Scenarios of multiparty data collection and the proposed FGNN framework
for predicting the network dynamics without compromising privacy. (a) Data
collection without privacy protection. Clients 1 and 2 are two local clients.
The rectangular boxes represent the time series data of the nodes with the
color of the box indicating the data at different timestamps. In this case,
each client has complete time series data of the nodes and complete information
about the network structure, so the information possessed by both clients is
transparent to each other, i.e., no privacy. (b) Each client collects only
partial timestamps of the time series data with incomplete knowledge about
the network structure (Scenario 1). (c) Each client collects only the time
series data from partial nodes but with complete knowledge about the network
structure (Scenario 2). (d) Proposed FGNN framework for predicting network
dynamics without compromising privacy.}
\label{fig:two scenes}
\end{figure}

To our knowledge, the increasingly critical issue of privacy has not been
addressed in the literature on data-based prediction of network dynamics. In
fact, a tacit assumption employed in the current literature on data-based
prediction of network dynamics is that the observed data are transparent and
available to all the observers, which include the network structure and the
time series data, as schematically illustrated in Fig.~\ref{fig:two scenes}(a).
If all the data are collected by a single client, privacy is not an issue.
However, in applications, practical limits such as the cost of observations
and the timeliness render necessary employing different clients to collect
the data~\cite{mooney2018big}. For example, disease-related data at different
times and/or in different regions are often collected by multiple parties,
which cannot be shared due to the requirements of data security and privacy
protection, leading to the emergence of the so-called
``data islands''~\cite{yu2020efficient,lenert2020balancing}. In the era of big
data analytics and machine learning, data have significant commercial,
security, and applied values. While the pertinent entities are able to better
accomplish their goals with more data, privacy protection puts a limit on
how much data any individual client is able to acquire. For network dynamics,
more time series data and more information about the network structure are
certainly beneficial to achieving higher accuracies in predicting the
dynamical evolution, but the data need to be collected by individual clients
for which privacy may be of great importance. A key question with practical
significance is how to coordinate the network structure and time series data
distributed in different organizations to improve the accuracy of dynamics
prediction, without compromising privacy.

Existing cooperative data-learning methods include secure multiparty
computing~\cite{atallah2001}, multitask learning~\cite{zhang2013}, and
federated learning~\cite{konevcny2016}. As a distributed machine learning
technology with privacy protection where data fusion can be achieved among the
parties without leaking their private data~\cite{yurochkin2019}, federated
learning has gained much interest in applications such as traffic flow
prediction~\cite{liu2020} and recommendation systems~\cite{wu2021}.
Federated learning has also been exploited for graph representation
learning~\cite{sattler2019,mei2019} and its downstream tasks such as node
classification, link prediction, and graph
classification~\cite{he2021,yang2019,chen2021}. The basic principle underlying
federated learning is that each client trains a local model with local data,
and a joint global model is generated by aggregating the parameters of the
local models~\cite{mcmahan2017}. Whether the federated learning framework can predict the network dynamics and how to use it to predict the network dynamics based on multiparty data have not been considered. To this end, we set out to develop a framework based on Federated Graph Neural Networks
(FGNNs) to predict the network dynamics from distributed time-series data
through jointly learning an optimal global dynamics model without exposing
the data of each party, i.e., without compromising privacy.

The problem of predicting the network dynamics without compromising privacy is
significantly more challenging than that of inferring the network structure,
for the following three reasons. First, different clients record only partial
time series data of (partial timestamps or partial nodes) of the network
dynamics. Second, it is necessary to learn the rules of the dynamical evolution
from the data. Third, the network data held by different clients are in general
not iid (independently and identically distributed). To make the forecasting problem with full privacy protection addressable
at the present, in this paper we focus on two data-collection scenarios. In
Scenario 1, each client collects only the partial time series data of all
nodes and the partial structure of networks, as illustrated in
Fig.~\ref{fig:two scenes}(b). In Scenario 2, each client has only the time
series data from a subset of nodes, but each client has complete knowledge
about the structure of the network, as shown in Fig.~\ref{fig:two scenes}(c).
We demonstrate that the
prediction performance of the FGNN framework far exceeds that of the local
models obtained by the individual local clients through local data. We also
show that the FGNN framework is capable of predicting the evolution of a
diverse array of discrete and continuous network dynamical processes. Overall,
the FGNN framework represents a powerful machine-learning based approach to
predicting the global network dynamics from only local data with guaranteed
privacy protection.

\section{The proposed FGNN framework} \label{sec:methods}

In this section, the architectures of the FGNN framework and its details are described.

\subsection{Basic principle of proposed FGNN framework} \label{subsec:models}

Let $G = \{V, E\}$ denote a given network, where $V=\{{v_i}|i=1,\cdots,n\}$
and $E = \{ {e_{ij}}|i,j = 1, \cdots ,n\}$ are the sets of nodes and
edges, respectively. The available time series are organized into a data
matrix $\mathcal{X}$, where each row corresponds to the time series of one
node and each column is associated with one timestamp of all nodes. The data
matrix $\mathcal{X}$ is generated by an unknown dynamical process $\cal D$ on
the underlying network (In this work, four discrete and three continuous dynamics are considered,  and detailed
description of the various dynamical processes are presented in Sec.~\ref{subsec:dyn} in Appendix).

Let ${X^t}$ be a vector that stores the states of all
nodes at time $t$ (i.e., a column of matrix $\mathcal{X}$), and the nodal
state evolution over time $t$ is given by
\begin{align} \label{eq:xt}
X^{t + 1} = {\cal D}{\rm{(}}{X^t},G{\rm{)}}.
\end{align}
Suppose there are $K$ clients and a client $k$ views the network as its own
local network structure $G_k$ and the available time series data is denoted as
$\mathcal{X}_k$ (i.e., a submatrix of $\mathcal{X}$). The time series data of
client $k$ are expressed as
$\mathcal{X}_k = {\rm{\{ }}X_k^1{\rm{,}}...{\rm{,}}X_k^{{T_k}}{\rm{\} }}$,
where ${T_k}$ is the length of the time series recorded by client $k$.
In Scenario 1, each client has partial timestamps of the whole time series
data $\mathcal{X}$, and $X_k^t \in {R^{n \times 1}}$ denotes the states of all
nodes at time $t$ by client $k$. In Scenario 2, the time series data held by
each client are at the same timestamp: ${T_k} = T,k = 1,2, \cdots ,K$. Let
$V_k$ be the nodal set whose time series data can be observed by client $k$,
where $X_k^t \in {R^{\left| {{V_k}} \right| \times 1}}$. The state of the node
$i$ at time $t$ is expressed as $x_k^t(i) = (X_k^t)_i$.

Each client $k$ can train a local machine learning model $M_k$ to generate the
network dynamics based on its own data, with a training parameter set denoted
as $\theta_k$. Because of the incomplete observation, the learning
capability of any local model in capturing the network dynamics is limited.
To overcome the limitation, we build up an FGNN framework by combining the
information of ${\rm{\{ }}{G_1}{\rm{,}}\cdots{\rm{,}}{G_K}{\rm{\} }}$ and
${\rm{\{ }}{\mathcal{X}_1}{\rm{,}}...{\rm{, }}{\mathcal{X}_K}{\rm{\} }}$ to
train a global model $M$ enabling us to better learn the dynamical
process and to accurately predict the future nodal states of nodes. In
particular, we have
\begin{align} \label{eq:D}
M{\rm{(}}{X^t},G,\theta {\rm{)}} \approx {\cal D}{\rm{(}}{X^t},G{\rm{)}},
\end{align}
where $\theta $ denotes the parameter set of the global model.

The overall framework of FGNN is illustrated in Fig.~\ref{fig:two scenes}(d), and it involves four main steps:

\textit{Step 1:}  The central unit initializes the model parameter set
$\theta^0$ and distributes them to each client.

\textit{Step 2:}  At the $t$th iteration, each client $k$ uses the new
parameter set $\theta^t$ to update its local model $M_k$. Each client $k$ takes
the current states $X_k^t$ as the input at time $t$ and outputs at the next
time step the nodal states
$\widehat {Y_k^t}={M_k}{\rm{(}}X_k^t,G_k,{\theta _k^t}{\rm{)}}$
based on its own local network structure $G_k$. The real states used for
training is $Y_t^k = X_{t + 1}^k$. To train the model, the loss function of
the local model is constructed by the real states and the output as
\begin{align} \label{eq:Loss}
{\cal L}{\rm{(}}\theta _k^t{\rm{)}} = \frac{1}{{\left| {{V_k}} \right|}}\sum\limits_{i \in {V_k}} L {\rm{( }}y_k^t{\rm{(}}i{\rm{)}},\widehat {y_k^t}{\rm{(}}i{\rm{))}} ,
\end{align}
where $L{\rm{( }}y_k^t{\rm{(}}i{\rm{)}},\widehat {y_k^t}{\rm{(}}i{\rm{))}}$ is
the loss function of node $i$ between the real state and the predicted output.
Once the loss function is defined, the parameter set $\theta_k^{t + 1}$ of the
local models can be updated by a standard back-propagation neural network model (see Sec.~\ref{subsec:FGNN} for further details).

\textit{Step 3:} The updated parameters of all local models are sent to the
central unit and the parameters of the global model $\theta$ is aggregated
by the weights ${\rm{\{}}{\theta _1}{\rm{,}}\cdots{\rm{,}}{\theta _K}{\rm{\}}}$
expressed as
\begin{align} \label{eq:theta}
{\theta ^{t + 1}} = \sum\limits_{k = 1}^K {{w_k}} \theta _k^{t + 1},
\end{align}
where ${{w_k}}$ is the aggregation weight measuring the quality of the data of
client $k$, and its details are given in Sec.~\ref{subsec:para}. The new updated parameter set
$\theta ^{t + 1}$ is sent back to each client.

\textit{Step 4:} Repeating \textit{steps 2-3} until the loss function converges
or reaches a given number of training times, and the global FGNN model is the
trained joint model.

\subsection{Local model and loss functions} \label{subsec:FGNN}
We use a three-layer neural network to construct the local model. To address that our framework can incorporate different GNN models, two widely used GNNs, i.e., the graph convolutional network (GCN) and
the graph attention network (GATN) are used in the hidden layer. The input
of the model is the nodal states data matrix $\mathcal{X}$. For discrete
dynamics, one-hot coding is used as the inputs. For continuous dynamics,
the continuous dynamical variables are taken directly as the inputs. The
states of the nodes are embedded into a $d$-dimensional feature space through
a linear layer ${f_1}: R^S \to {R^d}$, where $S$ denotes the dimension of the
inputs. The second layer utilizes GCN or GATN to aggregate the information
from the neighbors of a node. The model outputs the
prediction $\widehat Y$ through a linear layer.

The output of the discrete dynamics is the normalized probability vector
$\widehat{{P_i}}$ that the node $i$ belongs to different discrete states,
with the element ${\widehat p_{i,m}}$ being the prediction probability of
node $i$ in the $m$-th state. The state with the highest probability in the
vector is taken as the predicted state $\widehat y{\rm{(}}i{\rm{)}}$ of
node $i$. The cross entropy (CE) loss function is used for discrete dynamics,
which is defined as
\begin{align} \label{eq:theta_1}
{L_{CE}}{\rm{(}}{P_i},\widehat {{P_i}}{\rm{)}} =  - \sum\limits_m {{p_{i,m}}} \log {\widehat p_{i,m}},
\end{align}
where ${P_i}$ is the one-hot coding of the true state of node $i$ and
${p_{i,m}}$ is the $m$th element of the vector.

The output of the continuous dynamics $\widehat y$ is a one-dimensional
continuous variable. For continuous dynamics prediction, we use the mean
square error (MSE) loss function:
\begin{align} \label{eq:theta_2}
{L_{MSE}}{\rm{(}}y{\rm{(}}i{\rm{)}},\widehat y{\rm{(}}i{\rm{))}} = {{\rm{(}}y{\rm{(}}i{\rm{)}} - \widehat y{\rm{(}}i{\rm{))}}^2}.
\end{align}
The specific architectures of the neural networks for the discrete dynamics
and continuous dynamics are summarized in Table~\ref{tab7}.

\begin{table*}
\renewcommand\arraystretch{1.3}
\centering
\normalsize
\caption{Neural network architectures for discrete and continuous dynamics}
\setlength{\tabcolsep}{1.8mm}{
\begin{tabular}{lcc}
\hline
    &Discrete dynamics& Continuous dynamics\\\hline
    \multirow{3}{*}{Input layer}
  &One-hot (1, $S$)&Linear (1,32)\\
  &Linear ($S$, 32)&ReLU \\
  &ReLU & \\\hline
  \multirow{2}{*}{Hidden layer}&GCN(32,32) or GATN (32,32)&GCN(32,32) or GATN (32,32) \\
  &ReLU&ReLU \\\hline
  \multirow{2}{*}{Output layer}&Linear (32, $S$)&Linear (32,1) \\
  &Softmax &ReLU \\\hline
  Loss function&Cross-Entropy Loss&Mean-Squared Loss\\\hline
\end{tabular}}
\label{tab7}
\end{table*}

\subsection{Weighted aggregation of parameters} \label{subsec:para}


We introduce a unified method to evaluate the quality of the local data.
Assuming that the amount of time series data owned by each client $k$ is
$\left|D_k\right|$ and the amount of network structure data corresponds to the
number of edges in it, denoted by $\left|E_k\right|$, we set the aggregation
weight in model $k$ as
\begin{align} \label{eq:theta1}
{w_k} = \frac{1}{2}{\rm{(}}\frac{{\left| {{D_k}} \right|}}{{\left| {{D_1}} \right| +  \cdots  + \left| {{D_K}} \right|}} + \frac{{\left| {{E_k}} \right|}}{{\left| {{E_1}} \right| +  \cdots  + \left| {{E_K}} \right|}}{\rm{)}}.
\end{align}
For Scenario 1, each client $k$ has the time series data of all nodes at
different timestamps (i.e., $T_k$) and the partial structure of the network
($G_k$). The aggregation weight in Eq.~(\ref{eq:theta}) can be rewritten as
\begin{align} \label{eq:theta2}
{w_k} = \frac{1}{2}{\rm{(}}\frac{{{T_k}}}{{{T_1} +  \cdots  + {T_K}}} + \frac{{\left| {{E_k}} \right|}}{{\left| {{E_1}} \right| +  \cdots  + \left| {{E_K}} \right|}}{\rm{)}}.
\end{align}
For Scenario 2, all clients have the same network structure:
$G_k=G,~i=1,2,\cdots, K$, so it is not necessary to consider the data quality.
The length of time series data in all clients is the same (the whole length of
the original data), but each client only records the time series data on
a subset of nodes. Consequently, the amount of data in each client $k$ can be
simply denoted as $\left|V_k\right|$, yielding
\begin{align} \label{eq:theta3}
{w_k} = \frac{{\left| {{V_k}} \right|}}{{\left| {{V_1}} \right| +  \cdots  + \left| {{V_K}} \right|}}.
\end{align}

\subsection{Evaluation metrics}\label{sub:evaluation}
For discrete dynamics, we use the ACC index to measure the prediction accuracy,
defined as
\begin{align} \label{eq:theta_3}
ACC = \sum\limits_{i \in {V_T}} {\frac{{{\rm{I(}}y{\rm{(}}i{\rm{)}} = \widehat y{\rm{(}}i{\rm{))}}}}{{\left| {{V_T}} \right|}}} ,
\end{align}
where ${\rm{I(}} \cdot {\rm{)}}$ is an indicator function and
${\left| {{V_T}} \right|}$ is the number of nodes in the test set ${{V_T}}$. A larger value of ACC indicates a
higher prediction capability of the model.

For continuous dynamics, the differences between the true and predicted values
of the dynamical variables are taken to be the prediction error. We choose two
metrics to characterize the error: the mean square error (MSE) and the mean
absolute percentage error (MAPE). In particular, we use MAPE to quantify the
error for the mutualistic and gene dynamics. For CML dynamics, the values of
the nodal dynamical variables are in the unit interval, and
${y{\rm{(}}i{\rm{)}}}$ is in the denominator of the MAPE metric. The value of
MAPE can be large when ${y{\rm{(}}i{\rm{)}}}$ converges to zero, so we use
the MSE metric for CML dynamics, where $\sigma $ is defined as:
\begin{align} \label{eq:theta_4}
\sigma  = \left\{ \begin{array}{l}
	\frac{1}{{\left| {{V_T}} \right|}}\sum\limits_{i \in {V_T}} {\frac{{\left| {y{\rm{(}}i{\rm{)}} - \widehat y{\rm{(}}i{\rm{)}}} \right|}}{{y{\rm{(}}i{\rm{)}}}}}, \ \  \mbox{Mutualistic/Gene},\\
	\frac{1}{{\left| {{V_T}} \right|}}\sum\limits_{i \in {V_T}} {{{{\rm{(}}y{\rm{(}}i{\rm{)}} - \widehat y{\rm{(}}i{\rm{))}}}^2}}, \ \ \mbox{CML}.
\end{array} \right.
\end{align}

\section{Performance characterization and demonstration}
\label{sec:performance}

\subsection{Generation of training data}

First, the time series data of all nodes are generated by certain dynamics. If a dynamical
process leads to a steady state, it is not possible to uncover the network
dynamics. To ensure that sufficiently long time series data can be obtained,
the states of nodes are re-initialized after some steps of evolution so as
to prevent the states of nodes from entering any steady state. The states of
the first timestamps are stored in the data matrix $\mathcal{X}$ as the input
of the model, and the next timestamps are stored in $\mathcal{Y}$ as the real
states of nodes. For each client, the time series data are generated by
intercepting the original data. Specifically, in Scenario 1, each client
uses the data of all nodes in different timestamps as the training set. In
Scenario 2, each client records the whole length of the time series data on
some nodes only, and the states of the missing nodes in the client are set as
zero. This may lead to conflicts with the state of nodes for discrete dynamics.
Our solution is to assign continuous values to the discrete states when
training a discrete dynamics model in Scenario 2. By so doing, the model and
train method are the same as for continuous dynamics.
In addition, to reflect that each client only knows partial structure of the
original network in Scenario 1, we randomly remove some edges following a
uniform distribution.

\subsection{Simulation settings of FGNN}\label{subsec:simu}

To present our results in a concise and clear way, here in the main text we include
results with the GCN (results with GATN are deferred to Sec.~\ref{subsec:gat} in Appendix).
Numerical experiments are performed on three synthetic
networks (scale-free(BA), small-world(WS), and Erd\"os-Renyi (ER) random)
and six real-world networks: Word~\cite{word2006}, Celegans~\cite{cele1998}
(Cele), USAir~\cite{Usair2003}, Metabolic~\cite{meta2005} (Meta),
Email~\cite{Usair2003}, Tap~\cite{tap2013} (see Sec.~\ref{subsec:net} in
Appendix for the structural
information of these networks).

For dynamical processes on networks, we use
a diverse array of discrete and continuous dynamics to demonstrate the general
applicability of our FGNN framework. In particular, we test four types of
discrete dynamics (susceptible-infected-recovered (SIR) epidemic spreading
dynamics~\cite{sir1998}, susceptible-infected-susceptible (SIS)
dynamics~\cite{sis2015}, threshold dynamics~\cite{threshold2010}, and Kirman
dynamics~\cite{kirman1993}) and three types of continuous processes (i.e., gene
regulatory dynamics (Gene)~\cite{gene2015}, mutualistic interaction dynamics
among species in ecology (Mutualistic)~\cite{JHSLGHL:2018}, and coupled map
lattices (CML)~\cite{Kaneko:1985,KT:book,zhang2019}). the detailed
descriptions of these dynamical processes are summarized in Sec.~\ref{subsec:dyn} in Appendix.

In our simulations, the number of client is $K=3$, where each client trains
a local model individually based on the local data. The local models can be
conveniently chosen as the baseline models, which are termed as Local\_1,
Local\_2 and Local\_3 for the three clients, respectively. A centralized model
(termed as Center) is trained by using the full network structure and the full
time series data. The model Center serves only the purpose of comparison,
it has no privacy protection as it utilizes all information of the clients. Unless specified otherwise, the
number of iterations for federated aggregation in our FGNN framework is set
to be ten, and the learning rates of training the GCN and GATN models are
0.001 and 0.0001, respectively. The test set contains 20 time pairs ($t$, $t+1$) of data. To reduce the
statistical uncertainties, the values of the results are averaged over 20
realizations.

\subsection{Results for simulated network dynamics} \label{subsec:results}

\begin{table*} [ht!]
\renewcommand\arraystretch{1.1}
\centering
\normalsize
\caption{Five-step prediction performance of GCN model - basic component of FGNN, for different types of network dynamics.}
\setlength{\tabcolsep}{3mm}{
\begin{tabular}{lccccccc}
\hline
    &\multicolumn{4}{c}{Discrete dynamics (ACC)}&\multicolumn{3}{c}{Continuous dynamics ($\sigma $)}\\
    &SIR&SIS&Threshold&Kirman&Gene&Mutualistic&CML\\\hline
  $T+1$&0.87&0.85&0.80&0.92&0.672&1.168&0.025\\
  $T+2$&0.83&0.78&0.75&0.85&0.756&1.460&0.024\\
  $T+3$&0.78&0.75&0.73&0.83&0.820&1.647&0.024\\
  $T+4$&0.82&0.71&0.74&0.79&0.963&1.740&0.033\\
  $T+5$&0.80&0.70&0.72&0.81&1.001&1.718&0.030\\
  \hline
\end{tabular}}
\label{tab1}
\end{table*}

In our FGNN framework, a basic component is a centralized GNN. To demonstrate
that GNNs are capable of predicting the network dynamics, we use the complete
time series data and network structure to train the parameters in the GCN.
The length of the time series in discrete dynamics is set as $T=200$, and that
for continuous dynamics is $T=100$. In the trained GCN model, the nodal states
at time $T$ are then taken as the inputs with the states at time $T+1$ as the
prediction. Then, the predicted states at $T+1$ are inputs to the model to
yield the predicted states at $T+2$, and so on. Table~\ref{tab1} presents the
five-step prediction results of the GCN model on a scale-free network for time
from $T+1$ to $T+5$. It can be seen that, for the four types of discrete
dynamics, when the true states of nodes are inputted at $T$, the value of ACC
at $T+1$ is persistently larger than $80\%$ (for the Kirman dynamics, the value
is about $92\%$). Subsequently, when the predicted states are used as inputs,
the ACC values somewhat decrease but they are still above $70\%$ for the next
five time steps. For the continuous dynamics, Table~\ref{tab1} shows the error
$\sigma$ between the predicted and true state values for time from $T+1$ to
$T+5$.

The results from predicting the continuous dynamics indicate uniformly small
prediction errors, regardless of the specific types of processes. The results
in Table~\ref{tab1} thus demonstrate that the GCN model is capable of making
accurate short-term state prediction for different types of network dynamical
processes, paving the way for the model to be incorporated into our federated
learning framework in which the prediction task is accomplished by fusing data
from different clients with their privacy fully protected.

\begin{figure} [ht!]
\centering
\includegraphics[width=\linewidth]{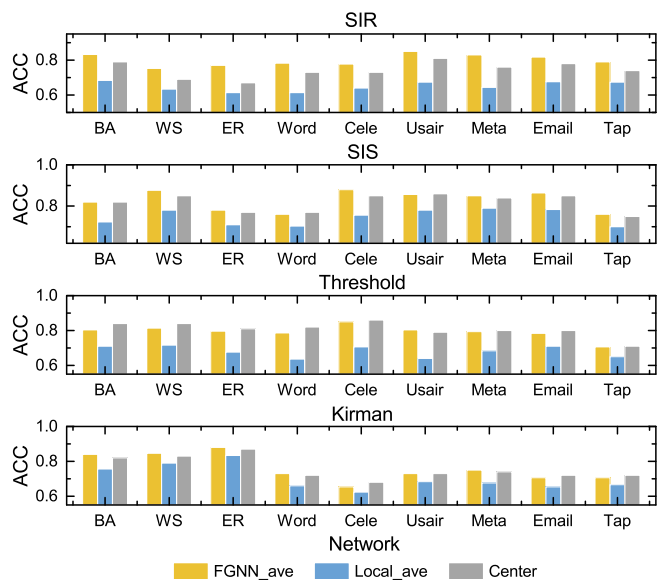}
\caption{Comparison of prediction performance of FGNN, local, and centralized
models for discrete dynamics for Scenario 1. Shown are the average values of
ACC from different models. The FGNN models significantly outperform the local
models and demonstrate a similar performance level to that of the centralized
model but with the desired advantage of full privacy protection.}
\label{fig:Scene1 discrete}
\end{figure}

\begin{figure} [ht!]
\centering
\includegraphics[width=\linewidth]{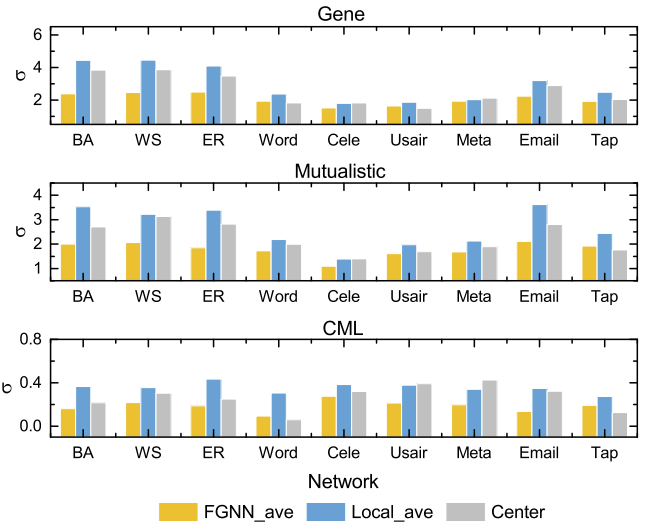}
\caption{Comparison of prediction performance of FGNN, local, and centralized
models for continuous dynamics for Scenario 1. Shown are the average errors
$\sigma$ for different models. As for the case of discrete dynamics in
Fig.~\ref{fig:Scene1 discrete}, the FGNN models significantly outperform the
local models and present a similar performance level to that of the centralized
model but without compromising privacy.}
\label{fig:Scene1 continuous}
\end{figure}

We can now present the prediction results of our FGNN framework. For Scenario
1 of data collection, each client $k$ has its own network structure $G_k$,
which is not shared with others, leading to three global prediction models
($K=3$), denoted as FGNN$\_k$=$M(\mathcal{X},\theta,G_k)$ for $k=1,2,3$.
For this case, the sampling
probabilities of edges in the three clients are set as 80\%, 60\% and 50\%,
respectively. All clients are assumed to have the same nodal set, and the
nodes without edges are treated as the isolated nodes. For the four types of
discrete dynamics, the lengths of the time series in the three clients are set
as 50, 30 and 20, respectively. For the three types of continuous dynamics,
the lengths of time series data in the three clients are set as 20, 15 and 15,
respectively. The three local models (Local$\_k$=$M_k(X_k,\theta_k, G_k)$) and the centralized
model (Center) serve as the baseline models for comparison. For clarity, here
we present the average results from the three FGNNs and three local models,
denoted as FGNN\_ave and Local\_ave, respectively.

Figure~\ref{fig:Scene1 discrete} shows, for data Scenario 1, the ACC values for
the FGNN, local and the centralized models in predicting the discrete dynamics
taking place on different networks. It can be seen that the FGNN models
persistently outperform the local models in term of the ACC values, with
similar or slightly better performance than that of the centralized model.
Considering that there is no privacy protection in the centralized model as
it requires complete time series data and complete information about the
network structure, our FGNN models are desired as it offers full privacy
protection. Figure~\ref{fig:Scene1 continuous} further compares the
performance of different models in predicting the continuous network dynamics
in terms of the measure $\sigma$, which again indicates that our FGNN framework
gives significantly better prediction results than those from the local
models and similar or slightly better results than the centralized model.
The results in Figs.~\ref{fig:Scene1 discrete} and \ref{fig:Scene1 continuous}
thus demonstrate that our FGNN framework can effectively combine multiple or
fuse local data information to train a better global network dynamic model
without disclosing the private data, regardless of the types of network
dynamics and structure.

\begin{table*}
\renewcommand\arraystretch{1.1}
\centering
\small
\caption{For Scenario 2, prediction performances of different models in term of MSE for four types of discrete dynamics for synthetic and
real-world networks. The best performance in each row is highlighted in bold.}
\setlength{\tabcolsep}{0.5mm}{
\begin{tabular}{lcccR{1cm}ccR{1cm}ccR{1cm}cc}
\hline
  Dynamic&\multicolumn{3}{c}{SIR}&\multicolumn{3}{R{1.8cm}}{SIS}&\multicolumn{3}{R{2.1cm}}{Threshold}&\multicolumn{3}{R{2cm}}{Kirman}\\
  Method&FGNN&Local\_ave&Center&FGNN&Local\_ave&Center&FGNN&Local\_ave&Center&FGNN&Local\_ave&Center\\\hline
  Scale-free &\textbf{1.371}&1.560&1.495   &\textbf{0.923}&1.533&1.216   &\textbf{0.654}&4.627 &1.531  &\textbf{0.992}&5.487 &4.402\\
  Small-world &\textbf{1.111}&1.619&1.530   &\textbf{0.818}&1.331&1.065  &\textbf{0.449}&0.767&0.473  &\textbf{0.432}&0.924&0.466\\
  ER random &\textbf{1.328}&1.768&1.654  &\textbf{1.024}&1.341&1.311  &\textbf{0.663}&1.972&1.279  &\textbf{0.722}&1.012&0.790\\
  Word&2.243&3.549&\textbf{2.229}  &\textbf{1.218}&2.380&1.651  &\textbf{0.416}&0.656&0.557  &\textbf{0.363}&0.386&0.386\\
  Cele&1.629&2.232&\textbf{1.430}  &\textbf{1.309}&2.193&1.688  &0.331&0.367&\textbf{0.330}  &0.526&0.458&\textbf{0.436}\\
  USAir&1.879&3.085&\textbf{1.716}  &1.291&1.538&\textbf{1.271}  &\textbf{0.465}&0.981&0.502  &\textbf{0.458}&0.612&0.514\\
  Meta&1.358&\textbf{1.220}&1.241  &1.276&1.645&\textbf{1.106}  &0.516&0.596&\textbf{0.507}  &0.860&1.815&\textbf{0.468}\\
  Email&\textbf{1.255}&2.042&1.666  &1.140&1.522&\textbf{1.058}  &\textbf{0.422}&0.535&0.514  &\textbf{0.463}&0.594&0.494\\
  Tap&\textbf{1.206}&1.215&1.370  &1.452&1.722&\textbf{1.437}  &\textbf{0.501}&0.878&0.779  &\textbf{0.633}&0.839&0.918\\\hline
\end{tabular}}
\label{tab2}
\end{table*}

\begin{table*}
\renewcommand\arraystretch{1.1}
\centering
\small
\caption{For Scenario 2, prediction performances of different models in term of $\sigma$ for three types of continuous dynamics for synthetic
and real-world networks. The best performance in each row is highlighted in
bold.}
\setlength{\tabcolsep}{0.5mm}{
\begin{tabular}{lcccR{1cm}ccR{1cm}cc}
\hline
  Dynamic&\multicolumn{3}{c}{Gene}&\multicolumn{3}{R{2.2cm}}{Mutualistic}&\multicolumn{3}{R{1.8cm}}{CML}\\
  Method&FGNN&Local\_ave&Center&FGNN&Local\_ave&Center&FGNN&Local\_ave&Center\\\hline
  Scale-free &\textbf{1.165}&1.432&1.406  &\textbf{1.245}&1.781&1.480  &0.529&0.839&\textbf{0.426}\\
  Small-world &\textbf{1.331}&1.714&1.648  &\textbf{1.815}&2.419&1.941  &0.358&0.990&0.356\\
  ER random &0.797&0.878&\textbf{0.766}  &\textbf{1.577}&1.847&1.616  &0.648&0.888&\textbf{0.395}\\
  Word&\textbf{0.800}&1.046&0.870  &\textbf{1.113}&1.275&1.223  &\textbf{0.538}&0.959&0.581\\
  Cele&0.889&1.448&\textbf{0.793}  &\textbf{1.574}&2.003&2.098  &\textbf{0.404}&0.878&0.416\\
  USAir&\textbf{1.230}&1.393&1.415  &\textbf{1.619}&2.192&2.240  &\textbf{0.481}&0.829&0.645\\
  Meta&1.402&1.323&\textbf{1.076}  &1.992&1.967&\textbf{1.719}  &\textbf{0.462}&0.702 &0.477\\
  Email&\textbf{1.161}&1.480&1.289  &\textbf{1.522}&2.152&1.886  &0.298&0.325&\textbf{0.281}\\
  Tap&\textbf{1.322}&1.540&1.587  &\textbf{2.372}&2.921&2.779  &\textbf{0.413}&1.376&1.279\\\hline
\end{tabular}}
\label{tab3}
\end{table*}

For data Scenario 2, different clients collect the time series data from a
subset of nodes but each client possesses the same global network structure, so
there is only one global federated learning model, denoted as
FGNN=$M(\mathcal{X},\theta,G)$. For this case, the length of the time
series for the four types of discrete dynamics is set to be 50, and the length of time series data for the three types of
continuous dynamics is set to be 20. In addition, the percentages of the nodes with data in the three
clients are 70\%, 80\% and 80\%, respectively. As we have mentioned in Sec.~\ref{subsec:simu}, for Scenario 2,  we treat the
discrete state as the continuous state to avoid the adverse consequence. Therefore, the MSE index is selected as the evaluation metric for the discrete network dynamics rather than the ACC index. Table~\ref{tab2} presents the prediction results of discrete network dynamics in term of MSE from different models.
It can be seen that, in most cases the prediction results of our FGNN framework
with privacy protection are better than those of the local models. For the
centralized model, in spite of its use of the complete data, the prediction
performance is not significantly better than that of our FGNN framework.
Table~\ref{tab3} illustrates the prediction results for the three types of
continuous dynamics from different models. Similar to the results in
Table~\ref{tab2}, our FGNN model can better predict the true nodal states than
the local models and, in some cases, even outperforms the centralized model.

\begin{figure} [ht!]
\centering
\includegraphics[width=\linewidth]{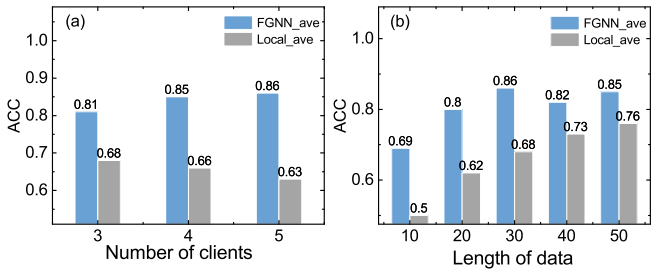}
\caption{Impacts of the number of clients (a) and the length of time series
(b) on the prediction performance for different models. The dynamical process
is of the SIR type, the network is of the scale-free type, and data collection
follows Scenario 1. In all cases, our FNGG model yields significantly better
results than the local models.}
\label{fig:mingan}
\end{figure}

To further demonstrate the robustness of our FGNN framework, we study the
impacts of the number of clients and the length of time series data on the
prediction performance. To be illustrative, we consider Scenario 1 and use
the SIR dynamics on scale-free networks. Figure~\ref{fig:mingan}(a) shows
the effect of the number of clients,
for $K=3$ (the sampling
probabilities of edges are 80\%, 60\%, 50\%, and the
length of the time series are 50, 30, 20), $K=4$ (the sampling
probabilities of edges are 80\%, 60\%, 50\%, 60\%, and the
length of the time series are 50, 30, 20, 15) and $K=5$ (the sampling
probabilities of edges are 80\%, 60\%, 50\%, 60\% and 50\%, and the
length of the time series are 50, 30, 20, 15, 10). As the number of clients
increases, the performance of the local models decreases, as each client
possesses less information, i.e., fewer time series data points and less
structural information about the network. In contrast, with more clients, our
FGNN models yields increasingly better performance - an intrinsic feature of
federated learning in general. Figure~\ref{fig:mingan}(b) presents the effect
of the time series length on the prediction performance for the case of $K=3$, where the length for
each client is the same for each simulation (10, 20, 30, 40 and 50). There is
no overlapping in the data for different clients. For example, if the total
length of the time series is $T=30$ and there are three clients, we set the
data length for each client to be 10. The sampling
probabilities of edges in the three clients are set as 80\%, 60\% and 50\%,
respectively. It can be seen from
Fig.~\ref{fig:mingan}(b) that the ACC values from our FGNN model are
persistently higher than those from the local models for all cases. For
shorter time series, the advantage of our FGNN model is more pronounced.
As expected, as the data length increases, the prediction performance of the
local models is improved as each client has more data to train the neural
network.

\subsection{Results for real-world network dynamics: predicting influenza evolution} \label{sec:influenza}

\begin{figure} [ht!]
\centering
\includegraphics[width=\linewidth]{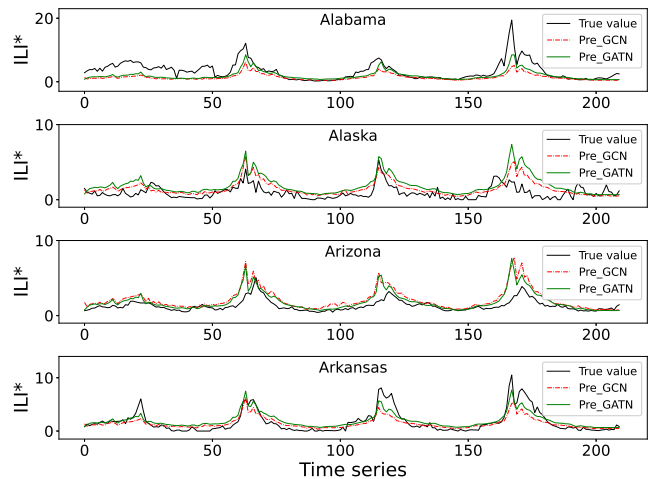}
\caption{Performance of two types of GNNs (without privacy protection) in
predicting the evolution of the influenza spreading in four States in the US.
The black, red dotted and green curves are the ground truth, and the predicted
ILI* values are GCN (Pre\_GCN) and GATN (Pre\_GATN).}
\label{fig:4 states}
\end{figure}

\begin{table*} [ht!]
\renewcommand\arraystretch{1.1}
\centering
\normalsize
\caption{MSE values of the global FGNN model and the baseline models under
data Scenario 1 in predicting the evolution of influenza in four States in
the US.}
\setlength{\tabcolsep}{1mm}{
\begin{tabular}{lccccccc}
\hline
  MSE&FGNN\_1&FGNN\_2&FGNN\_3&Local\_1&Local\_2&Local\_3&Center\\
  GCN&1.591&1.642&1.570&1.678&1.756&1.741&1.594\\
  GATN&0.685&0.814&0.822&1.413&1.394&1.506&1.271\\\hline
\end{tabular}}
\label{tab5}
\end{table*}

\begin{table} [ht!]
\renewcommand\arraystretch{1.1}
\centering
\normalsize
\caption{MSE values from the global FGNN and the baseline models under data
Scenario 2 in predicting the evolution of influenza in four States in
the US.}
\setlength{\tabcolsep}{1mm}{
\begin{tabular}{lccccc}
\hline
  MSE&FGNN&Local\_1&Local\_2&Local\_3&Center\\
  GCN&1.600 &1.622&2.584 &2.499 &1.852 \\
  GATN&1.227&1.646&1.352 &1.808&1.684 \\\hline
\end{tabular}}
\label{tab6}
\end{table}

We further apply our FGNN framework for predicting the dynamical states to a real-world
problem: outbreak of influenza. The time series are from the US weekly
influenza-like illness (ILI) data base~\cite{ili.org} for the five-year time
period from the 40th week of 2011 to the 39th week of 2016, which records
the weekly number of ILI-related visits to all public health and clinical
laboratories in the country by the CDC (Center for Disease Control). For data
pre-processing, we normalize the weekly ILI-related visits in each State to
calculate the ILI ratio, i.e., the percentage of the patients with influenza
among all visiting patients in each State~\cite{pei2018}. We then use the data
values in which the percentage of the ILI ratio is removed to represent the
state of the underlying dynamical system, record as ILI*. The underlying network supporting the influenza spreading across the
different States in the US is identified by taking advantage of the commuting
traffic data in different States in 2015 to generate a population commuting
network among the States~\cite{commute.org}. The raw data are available
from the website of the USA Bureau of Statistics, and the commuting data
among the States are obtained by aggregating the raw data
(A detailed description
of data processing and the data settings are provided in Sec.~\ref{subsec:ili} in Appendix).

As for the various cases of synthetic dynamic data treated in the preceding
subsection, the first step is to validate the effectiveness of the GNNs. For
this purpose, we use the influenza data at the \emph{previous time steps} to
train the GNNs. The trained GNNs and the ILI* of nodes (i.e., states) at the
\emph{current time step} are the inputs for predicting the trend of the
influenza at the \emph{next time step}. Figure~\ref{fig:4 states} shows the
true and the predicted values in term of ILI* for the first four States
in the alphabetic list of the US States. It can be seen that the predicted
values fit well the real evolution of the influenza in these States, and the
peaks of the influenza at several representative time points (i.e., outbreaks)
can also be predicted. The results in
Fig.~\ref{fig:4 states} thus demonstrate that the
predicted dynamics can fit the evolution of the influenza to a satisfactory
extent.

Having demonstrated the performance of the two GNNs, we can proceed to test the
predictive power of our FGNN framework and that of the baseline methods with
respect to the two data scenarios. The number of clients is
set to be $K=3$. For data Scenario 1, partial timestamps of all nodes are
collected and the information about the structure of the commuting network
is incomplete. For Scenario 2, the time series data of partial nodes are used.
Table~\ref{tab5} shows, for Scenario 1, the MSE values associated with the
FGNN predictions with the GCN and GATN models, the local models and the
centralized model. Regardless of whether the global model is GCN or GATN, our
FGNN models of the three clients (FGNN\_1, FGNN\_2, FGNN\_3) can achieve
better prediction performance than those of the local models (Local\_1,
Local\_2, Local\_3) and the centralized model (Center). Table~\ref{tab6}
summarizes the MSE values for the FGNN and baseline methods for Scenario 2.
It can be seen that our global FGNN model also yields significantly better
prediction results than those of the baseline models. For both data scenarios,
Tables~\ref{tab5} and \ref{tab6} unveil a surprising result: the prediction
performance of the FGNN model exceeds that of the centralized model that
uses the complete data (and so does not offer any data privacy protection).
A plausible reason is that our FGNN framework relies an iterative process
and so it has more training time and yields a better learning model.

\section{Discussion}\label{sec:discussion}

In our modern time, time series data generated from network dynamical
processes and the underlying structure of the network are typically owned by
different parties. There are two possible data distribution scenarios that
are amenable to simulations and analysis at the present: Scenario 1 in which
different clients have the data timestamps of all nodes but each client has
only partial information about the structure of the network, and Scenario 2 in
which different clients have the data from only a subset of nodes in the
network but each client has the full knowledge of the network structure. To
combine multi-party data to better predict network dynamics while protecting
privacy of the clients is a problem of great importance and interest. We have
proposed an FGNN framework to address this problem. The essence of our
prediction framework is to combine the data from different clients to jointly
train a high-quality global model without disclosing any private data. The
framework is compatible with existing GNNs to learn network dynamics and
predict the states of nodes into the near future. We have used two
classical GNN models (GCN and GATN) to demonstrate the power and effectiveness
of our FGNN framework through extensive numerical simulations of a good number
of synthetic and real-world networks as well as a variety of discrete and
continuous network dynamics. Our systematic comparison of the performances of
our global FGNN model, baseline local and centralized models under different
conditions reveals that the global model learned through the FGNN framework
has superior accuracy in predicting the dynamical evolution of the network and
is thus capable of better capturing the complex relationship between dynamics
and network structure. The impacts of the number of clients and the length of
time series data on the dynamics prediction have been investigated, revealing
the general applicability and robustness of our FGNN framework.

A practically significant contribution of our work is the demonstration of
successful prediction of the evolutionary trend of influenza in the US by
our articulated FGNN framework without compromising privacy. In particular,
any State in the US is regarded as a node in the network which we have
reconstructed using the State-crossing commuting data, and the short-term
dynamical evolution of the influenza spreading dynamics on this network has
been predicted. Our results indicate that, not only is the FGNN framework
capable of reproducing the actual time evolution of the influenza in the
States, but the outbreaks (corresponding to peaks of the dynamical evolution)
can also be faithfully predicted.

The two data scenarios treated in this paper are somewhat specific. A more
general scenario is that each client has partial time series data from a subset
of nodes and incomplete information about the network structure. To predict
the dynamical evolution of the network without compromising data privacy under
this general scenario is an open question warranting further investigation.

\acknowledgments

The authors are grateful to Drs.~Wen Hu and Xiao Ding for helpful discussions.
We also thank Dr.~Sen Pei for sharing the website of the influenza data in the
USA. This work is supported by National Natural Science Foundation of China
(Grant No.~61973001) and by The University Synergy Innovation Program of Anhui
Province (Grant No.~GXXT-2021-032). The work at Arizona State University was
supported by AFOSR under Grant No.~FA9550-21-1-0438.

\section*{Appendix}
\subsection{Structural information of networks } \label{subsec:net}

Experiments are performed on three synthetic networks and six real-world
networks. The sizes of the synthetic networks are 100 nodes. In particular,
the scale-free (BA) networks are generated with $m=2$, where $m$ is the number of
edges connecting to the existing nodes at each preferential attachment step.
The small-world (WS) network are generated with the average degree
$\langle k \rangle = 4$ and reconnection probability $p =0.3$. The
Erd\"os-Renyi (ER) random networks are generated with the edge connection
probability $p = 0.08$. The structural information of the six real-world
networks is summarized in Table~\ref{tabS1}, where $n$ and $M$ are the numbers
of nodes and edges of the network, respectively, $CC$ is the clustering
coefficient, $H=\langle k^2\rangle/\langle k\rangle^2$ is the network
heterogeneity, and $\langle k^2 \rangle $ is the second moment of the degree
distribution.

\begin{table}
\renewcommand\arraystretch{1.1}
\centering
\normalsize
\caption{Basic structural information of six real networks.}
\setlength{\tabcolsep}{1.5mm}{
\begin{tabular}{lccccc}
\hline
  Network&$n$&$M$&$\left\langle k \right\rangle $&$CC$&$H$\\
  Word&112&425&7.589&0.173&1.815\\
  Celegans&297&2148&14.465&0.292&1.801\\
  USAir&332&2126&12.807&0.749&3.464\\
  Metabolic&453&2025&8.940&0.647&4.485\\
  Email&1133&5451&9.622&0.220&1.942\\
  Tap&1373&6833&9.953&0.529&
  1.644\\\hline
\end{tabular}}
\label{tabS1}
\end{table}

\subsection{Seven types of network dynamics} \label{subsec:dyn}
We describe the four types of discrete and three types of continuous networked
dynamical processes used in our study, in our paper, which all have been well
studied to gain insights into a variety of network phenomena in natural or
social sciences.

{\em SIR dynamic model.} In the SIR model~\cite{sir1998}, at any time a node
can be in one of three states: susceptible ($S$), infectious ($I$) and
recovery ($R$). The $I$-state node can infect its $S$-state neighbors with
the infection rate $\lambda$. If an infection event is successful, the infected
node will change its state from $S$ to $I$; Otherwise it will remain in the
$S$ state. The infected nodes will recover to the $R$ state with the recovery
rate $\mu$, and the $R$-state nodes will not be infected again. In our study,
the parameter values are set as $\lambda =0.2$ and $\mu  = 0.1$. The states of
all nodes are randomly initialized after every ten time steps.

{\em SIS dynamic model.} In the SIS model~\cite{sis2015}, there are two
distinct dynamical states only: $S$ and $I$. An $I$-state node can infect
its susceptible neighbors with the infection rate $\lambda$ and recover to the
$S$ state with the recovery rate $\mu$. In our study, the parameter values are
set as $\lambda =0.2$ and $\mu  = 0.1$. The states of all nodes are randomly
initialized after every ten time steps.

{\em Threshold dynamic model.}
In the threshold model~\cite{threshold2010}, nodes can be in one of the two
states: inactive (0) and active (1). An inactive node is activated when the
fraction of its active neighbors is greater than an activation threshold of
the node, and an active node will be not restored again. In our study, the
activation threshold for all node is set to be 0.5, and the states of all
nodes are re-initialized after every five time steps.

{\it Kirman dynamic model.}
In the Kirman model~\cite{kirman1993}, a node can be in one of the two states:
0 and 1. The transition between the two states is based upon two transfer
functions. In particular, the transfer function of the node from 0 to 1 is
given by ${c_1} + d{m_1}$, and that from 1 to 0 is
${c_2} + d{\rm{(}}k - {m_1}{\rm{)}}$, where ${c_1}$ and ${c_2}$ quantify the
individual behavior of the node that is independent of neighbors' state, and
$k$ is the nodal degree, $d$ describes the probability of node replicating a
neighbor's state, and ${m_1}$ is the number of 1-state neighbors. In our study,
the parameters are set as ${c_1} = 0.1$, ${c_2} = 0.1$ and $d = 0.08$.

The continuous networked dynamics employed in our study are one-dimensional
with the state variable ${x_i(t)} \in {R}$ for node $i$ at time $t$. The
states of all nodes at time $t$ can be represented by the vector
$X{\rm{(}}t{\rm{)}} = {\rm{[}}{x_1}{\rm{(}}t{\rm{),}} \cdots ,{x_n}{\rm{(}}t{\rm{)]^T}} \in {R^{n}}$. The differential equations governing the three types of
continuous dynamics in our study are as follows.

{\em Gene regulatory dynamics (Gene).}
The gene regulation dynamics are expressed by the Michaelis Menten
equation~\cite{gene2015}:
\begin{align} \label{eq:transition_1}
\frac{{d{x_i}{\rm{(}}t{\rm{)}}}}{{dt}} =  - {u_i}{x_i(t)} + \sum\limits_{j = 1}^n {{A_{ij}}} \frac{{{{{\rm{(}}{x_j(t)}{\rm{)}}}^h}}}{{{{{\rm{(}}{x_j(t)}{\rm{)}}}^h} + 1}},
\end{align}
where the first item controls the decay of the current node state, ${u_i}$
is the decay rate, and the second term captures gene activation characterized
by the Hill coefficient $h$. We use ${u_i}=1$ and $h=2$. The states of all
nodes are re-initialized after 50 time steps.

{\em Mutualistic interaction dynamics (Mutualistic).}
The differential equations governing the evolution of mutualism in ecological
systems~\cite{mut2016,JHSLGHL:2018} are
\begin{align}\label{eq:transition_2}
\begin{array}{c}
\frac{{d{x_i}(t)}}{{dt}} = {u_i} + {x_i}(t)(1 - \frac{{{x_i}(t)}}{{{l_i}}})(\frac{{{x_i}(t)}}{{{z_i}}} - 1)\\
 + \sum\limits_{j = 1}^N {{A_{ij}}\frac{{{x_i}(t){x_j}(t)}}{{{\alpha _i} + {\beta _i}{x_i}(t) + {\gamma _i}{x_j}(t)}}}
\end{array} ,
\end{align}
where the first item $u_i$ represents the number of migrating individuals of
species $i$ from the adjacent habitat, the second item ${{l_i}}$ describes the
carrying capacity of the system growth, and ${{z_i}}$ is the cold-start
threshold. When the abundance of species $i$ is low (i.e., ${x_i(t)} < {z_i}$),
the system is characterized by a negative growth. The third item in
Eq.~\eqref{eq:transition_2} describes the interactions among the species, which
take place on a network described by the adjacency matrix $\mathcal{A}$. In our
study, we use the parameter values ${u_i}=0.1$, ${{l_i}}=5$, ${{z_i}}=1$,
${\alpha _i}=5$, ${\beta _i}=0.9$, and ${\gamma _i}=0.1$. The states of all
nodes are re-initialized after each 50 time steps.

{\it Coupled map lattices (CML).}
In a coupled mapping lattice, the continuous state variables are updated at
discrete time~\cite{Kaneko:1985,KT:book,zhang2019} according to
\begin{align} \label{eq:transition_3}
{x_i}{\rm{(}}t + 1{\rm{)}} = {\rm{(}}1 - s{\rm{)}}f{\rm{(}}{x_i}{\rm{(}}t{\rm{))}} + \frac{s}{k_i}\sum\limits_{j = 1}^N {{A_{ij}}f{\rm{(}}{x_j}{\rm{(}}t{\rm{))}}} ,
\end{align}
where $s$ is the coupling parameter (the system degenerates into a set of
independent mapping functions for $s = 0$) and $k_i$ is the degree of node
$i$. In our study, we use the mapping function
$f{\rm{(}}x{\rm{)}} = \lambda x{\rm{(}}1 - x{\rm{)}}$ for the parameters
$\lambda = 3.5$ and $s=0.2$. The states of all nodes are re-initialized after
each 50 time steps.

\subsection{GATN simulation results}\label{subsec:gat}
\begin{figure}
\centering
\includegraphics[width=\linewidth]{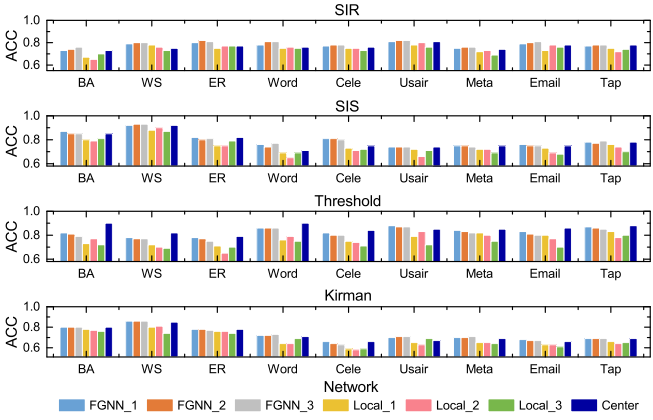}
\caption{FGNN/GATN model performance. Shown are the values of ACC from the
global models and the baseline methods for the four types of discrete dynamical
processes under data Scenario 1.}
\label{figS:scene1 disgat}
\end{figure}

\begin{figure}
\centering
\includegraphics[width=\linewidth]{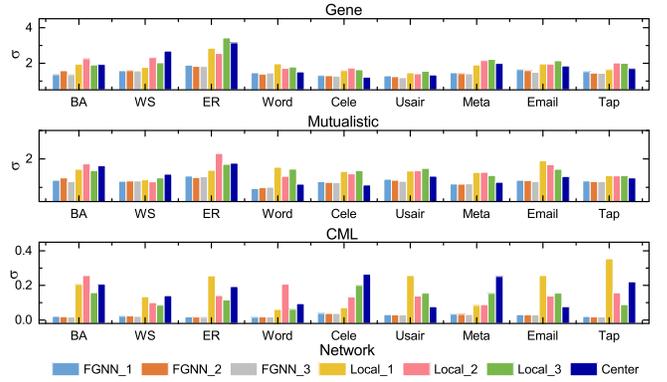}
\caption{FGNN/GATN model performance. Shown are the prediction error $\sigma$
from the global models and the baseline methods for the three types of
continuous dynamical processes under data Scenario 1.}
\label{figS:scene1 consgat}
\end{figure}

We present the detailed simulation results of the GATN-based FGNN model and the
local model of each client. Table~\ref{tabS5} verifies the good performance of
the GATN model in predicting the four discrete dynamics ($T=200$ for GATN
training) and the three continuous dynamics ($T=100$ for GATN training) with
the complete time series data and the complete network structure (for
scale-free networks). Figure~\ref{figS:scene1 disgat} shows the values of ACC
from the FGNN global models and the baseline methods for the four types of
discrete dynamics under data Scenario 1.  Figure~\ref{figS:scene1 consgat} shows the prediction
results for the three types of continuous dynamics, which also suggest that
the FGNN global models have lower prediction errors than those of the local
models. For data Scenario 2, the results in Table~\ref{tabS6} demonstrate that the
performance of our FGNN method in predicting the four types of discrete
dynamics is persistently better than that of the local models. Table~\ref{tabS7} shows the results of different methods in predicting the three types of continuous dynamics. In most cases, the
prediction results of our FGNN method are closer to the real values than
those from the local models.

\begin{table*}[h]
\renewcommand\arraystretch{1.1}
\normalsize
\caption{Five-step dynamics prediction results of GATN model based on complete
time series data and complete network structure.}
\setlength{\tabcolsep}{3mm}{
\begin{tabular}{lccccccc}
\hline
    &\multicolumn{4}{c}{Discrete dynamics(ACC)}&\multicolumn{3}{c}{Continuous dynamics($\sigma $)}\\
    &SIR&SIS&Threshold&Kirman&Gene&Mutualistic&CML\\\hline
  $T+1$&0.85&0.86&0.89&0.84&0.598&0.958&0.017 \\
  $T+2$&0.73&0.80&0.84&0.81&0.602&1.086&0.021 \\
  $T+3$&0.81&0.75&0.81&0.82&0.609&1.276&0.024 \\
  $T+4$&0.82&0.74&0.74&0.83&0.724&1.512&0.027 \\
  $T+5$&0.80&0.74&0.72&0.85&0.822&1.601&0.028 \\
  \hline
\end{tabular}}
\label{tabS5}
\end{table*}

\begin{table*}[h]
\renewcommand\arraystretch{1.1}
\centering
\tiny
\caption{Values of MSE from the global and baseline models for the four types
of discrete dynamics for FGNN/GATN under data Scenario 2.}
\setlength{\tabcolsep}{0.3mm}{
\begin{tabular}{lcccccR{0.8cm}ccccR{0.8cm}ccccR{0.8cm}cccc}
\hline
  Dynamic&\multicolumn{5}{c}{SIR}&\multicolumn{5}{R{2.4cm}}{SIS}&\multicolumn{5}{R{2.7cm}}{Threshold}&\multicolumn{5}{R{2.6cm}}{Kirman}\\
  Method&FGNN&Local\_1&Local\_2&Local\_3&Center&FGNN&Local\_1&Local\_2&Local\_3&Center&FGNN&Local\_1&Local\_2&Local\_3&Center&FGNN&Local\_1&Local\_2&Local\_3&Center\\\hline
  Scale-free&\textbf{0.843}&0.846&0.862&0.893&0.865  &\textbf{0.177}&0.286&0.484&0.533&0.261  &\textbf{0.413}&0.477&0.456&0.418&0.427  &\textbf{0.249}&0.435&0.269&0.255&0.259\\
  Small-world&\textbf{0.897}&1.022&1.034&1.046&1.220  &0.146&0.170&0.144&0.201&\textbf{0.136}  &\textbf{0.245}&\textbf{0.245}&0.282&0.375&0.311  &\textbf{0.259}&0.323&0.285&0.315&0.342\\
  ER random&0.649&\textbf{0.638}&0.696&0.674&0.655  &0.242&0.310&0.253&0.237&\textbf{0.211}  &\textbf{0.224}&0.245&0.266&0.273&0.302  &\textbf{0.242}&0.298&0.277&0.327&0.265\\
  Word&\textbf{0.699}&0.799&0.917&0.763&0.791  &\textbf{0.249}&0.255&0.269&0.494&\textbf{0.249}  &0.295&0.538&\textbf{0.268}&0.442&0.341  &\textbf{0.251}&0.277&0.265&0.265&0.271\\
  Cele&0.496&0.512&0.511&0.498&\textbf{0.478}  &0.270&0.282&0.278&0.272&\textbf{0.257}  &\textbf{0.301}&0.346&0.387&0.347&0.316  &\textbf{0.254}&0.309&0.307&0.266&0.317\\
  USAir&\textbf{0.699}&0.744&0.820&0.747&0.716  &\textbf{0.246}&0.249&0.347&0.363&0.252  &\textbf{0.401}&0.474&0.572&0.456&0.419  &0.251&0.266&0.265&0.433&\textbf{0.250}\\
  Meta&\textbf{0.681}&0.793&0.813&0.695&0.735  &\textbf{0.232}&0.254&0.248&0.29&0.250  &0.311&0.335&\textbf{0.286}&0.540&0.290  &\textbf{0.251}&0.289&0.314&0.317&0.283\\
  Email&\textbf{0.696}&0.750&0.765&0.731&0.771  &\textbf{0.251}&0.290&0.298&0.286&0.284  &\textbf{0.257}&0.446&0.301&0.350&0.759  &\textbf{0.251}&0.399&0.269&0.279&0.290\\
  Tap&\textbf{0.632}&0.669&0.675&0.662&0.701  &0.253&0.322&\textbf{0.248}&0.294&0.254  &0.250&0.305&0.262&0.284&\textbf{0.245}  &\textbf{0.246}&0.316&0.321&0.257&0.347\\\hline
\end{tabular}}
\label{tabS6}
\end{table*}

\begin{table*}[h]
\renewcommand\arraystretch{1.1}
\centering
\scriptsize
\caption{Values of prediction error from the global and baseline models for the three types
of continuous dynamics for FGNN/GATN under data Scenario 2.}
\setlength{\tabcolsep}{0.5mm}{
\begin{tabular}{lcccccR{1cm}ccccR{1cm}cccc}
\hline
  Dynamic&\multicolumn{5}{c}{Gene}&\multicolumn{5}{R{2.8cm}}{Mutualistic}&\multicolumn{5}{R{2.4cm}}{CML}\\
  Method&FGNN&Local\_1&Local\_2&Local\_3&Center&FGNN&Local\_1&Local\_2&Local\_3&Center&FGNN&Local\_1&Local\_2&Local\_3&Center\\\hline
  Scale-free&1.160&1.163&1.159&1.627&\textbf{1.122}  &1.741&1.834&\textbf{1.437}&1.733&1.636  &\textbf{0.123}&0.199&0.162&0.245&0.192\\
  Small-world&1.258&1.351&1.255&\textbf{1.120}&1.139  &\textbf{1.141}&1.295&1.382&1.184&1.264  &\textbf{0.156}&0.237&0.213&0.217&0.313\\
  ER random&0.774&0.823&0.788&0.811&\textbf{0.743}  &1.349&1.341&1.372&1.177&\textbf{1.308}  &0.191&0.354&0.249&0.174&\textbf{0.101}\\
  Word&\textbf{0.774}&1.019&0.916&1.201&0.884  &1.655&1.624&1.646&1.754&\textbf{1.442}  &0.229&0.197&0.235&0.287&\textbf{0.146}\\
  Cele&\textbf{0.686}&0.844&0.857&1.035&0.778  &\textbf{1.005}&1.202&1.250&1.580&1.316  &0.176&0.169&0.200&\textbf{0.161}&0.172\\
  USAir&\textbf{0.891}&1.018&1.009&1.097&0.945  &\textbf{1.165}&1.420&1.341&1.395&1.361  &\textbf{0.146}&0.155&0.198&0.170&0.148\\
  Meta&\textbf{0.990}&1.080&1.072&1.180&1.006  &\textbf{1.245}&1.394&1.407&1.490&1.481  &0.151&0.267&0.191&0.183&\textbf{0.088}\\
  Email&1.261&1.044&1.086&1.212&\textbf{1.055}  &\textbf{1.310}&1.626&1.409&1.583&1.489  &0.085&0.123&0.168&0.067&\textbf{0.053}\\
  Tap&0.955&1.040&0.969&1.038&\textbf{0.908}  &\textbf{1.207}&1.310&1.411&1.317&1.234 &\textbf{0.093}&0.176&0.137&0.122&0.109\\\hline
\end{tabular}}
\label{tabS7}
\end{table*}

\subsection{Influenza data}\label{subsec:ili}
\subsubsection{Description of influenza data}

\begin{figure}
\centering
\includegraphics[width=\linewidth]{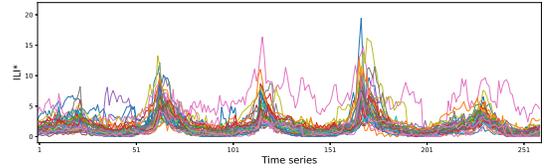}
\caption{Real ILI* values in the 37 States from the 40th week of 2011 to the
39th week of 2016.}
\label{figS:37 US}
\end{figure}

The influenza-like illness (ILI) data used in our study are from the US public
health and clinical laboratories~\cite{ili.org}, which were collected weekly
from the 40th week of 2011 to the 39th week of 2016. We standardize the weekly
ILI data by using the number of visiting patients in each State in the US to
calculate the ILI ratio~\cite{pei2018}. In order to ensure a sufficient sample
size in the data, we assume that the States whose average weekly ILI ratio is
less than 1\% are not representative. Accordingly, 14 States are removed. The
remaining 37 States are listed in Table~\ref{tabS2}. Because the small value
of ILI ratio (about 2$\%$) can lead to errors, we multiply the ILI ratio
by the factor of 100 (labelled as ILI*) and use the resulting values the
dynamic data. The curves of ILI* for the 37 States are shown in
Fig.~\ref{figS:37 US}.

\begin{table*}[h]
\renewcommand\arraystretch{1.1}
\centering
\small
\caption{Names of the 37 US States used in our study.}
\setlength{\tabcolsep}{1.8mm}{
\begin{tabular}{lcccccc}
\hline
  Alabama&Alaska&Arizona&Arkansas&California&Connecticut&Columbia\\
  Georgia&Hawaii&Idaho&Illinois&Indiana&Kansas&Louisiana\\
  Maryland&Massachusetts&Michigan&Minnesota&Mississippi&Missouri&Nebraska\\
  Nevada&New Jersey&New Mexico&New York&North Carolina&North Dakota&Oklahoma\\
  Pennsylvania&South Dakota&Tennessee&Texas&Utah&Vermont&Virginia\\
  West Virginia&Wisconsin& & & & &\\\hline
\end{tabular}}
\label{tabS2}
\end{table*}

\begin{figure}
\centering
\includegraphics[width=0.8\linewidth]{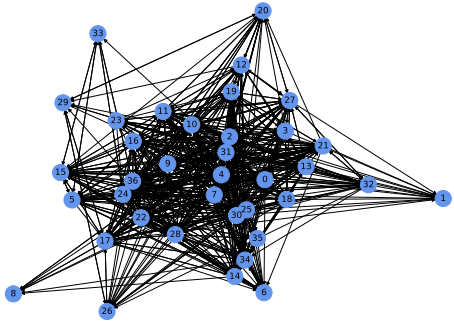}
\caption{The directed and weighted network constructed from the commuting data
among the 37 US States.}
\label{figS:graph}
\end{figure}

Population commuting between the States is the key to spreading dynamics. We
use the commuting data in the USA by residential geography to construct the
commuting network. The commuting data between the USA cities are from the 2015
census report~\cite{commute.org}. The State level commuting data are calculated
by aggregating the city data, leading to a State-level commuting network for
the 37 States. The network is directed and weighted, where the nodes represent
the States and the edges are determined by the commuting data among the States.
To prevent the network from being too dense, we use the criterion that, if the
number of commuters between the two states is less than 100, the corresponding
edge is removed. The resulting network has 763 directed edges. Finally, we use
the min-max normalization procedure~\cite{patro2015} to obtain the weights of
the edges by dividing the number of commuters by the maximum number of
commuters. The constructed directed and weighted network can be visualized,
as shown in Fig.~\ref{figS:graph}, where the thickness of the directed edges
denotes the weight.

\subsubsection{Simulation settings of influenza data}

Under Scenario 1, the first 100 data points are selected from a total of 260
ILI* data points. The numbers of data assigned to the three clients are 50,
30 and 20, respectively, and the network structures for the three clients are
generated by sampling 90\%, 80\% and 70\% of edges from the commuting network.
The learning rates of the FGNN/GCN and FGNN/GATN models are set to be 0.0001
and 0.001, respectively. Under data Scenario 2, the first 50 data points are
selected from a total of 260 ILI* data points. The percentages of nodes
with the influenza data for the three clients are 80\%, 70\% and 60\%,
respectively. The learning rates of the GCN and GATN models are set as 0.01
and 0.0001, respectively.

%

%

\end{document}